\documentclass[preprint]{ptephy_v2}

\usepackage{hyperref}
\usepackage{subcaption}

\usepackage{graphics}
\usepackage{dcolumn}
\usepackage{bm}
\usepackage[mathlines]{lineno}
\usepackage{upgreek}  
\usepackage{multirow}  
\usepackage{xcolor}
\usepackage{placeins}           
\usepackage{hyperref}
\usepackage{url}
\usepackage[mathlines]{lineno}

\begin{document}
\title{Measurement of the Quantum Efficiency of Electrode Materials for VUV Photons in Liquid Xenon}

\author[1,2,*]{S. Kazama}
\author[2]{N. Aoyama}
\author[1,2]{Y. Itow\thanks{Present Address: Institute for Cosmic Ray Research, University of Tokyo, Kashiwa, Chiba, Japan}}
\author[1]{M. Kobayashi}

\affil[1]{Kobayashi-Maskawa Institute for the Origin of Particles and the Universe, Nagoya University, Nagoya, Aichi, 464-8601, Japan}
\affil[2]{Institute for Space-Earth Environmental Research, Nagoya University, Nagoya, Aichi, 464-8601, Japan \email{kazama@nagoya-u.jp}}


\begin{abstract}%
Light dark matter searches using ionization signals in dual-phase liquid xenon (LXe) time projection chambers (TPCs) are limited by low-energy ionization backgrounds, including those from the photoelectric effect on the electrodes. To address this, we measured the quantum efficiency (QE) of various electrode materials for vacuum ultraviolet (VUV) photons in LXe, including platinum (Pt), stainless steel (SUS304), and magnesium fluoride (MgF$_{2}$)-coated aluminum (Al).  
Our results show that MgF$_{2}$-coated Al exhibits the lowest QE among the tested materials. The QE for VUV photons with a mean wavelength of 179.5\,nm was measured to be $(7.2 \pm 2.3) \times 10^{-5}$, corresponding to a reduction in QE by a factor of 4.4 compared to SUS304, a commonly used electrode material in direct dark matter experiments with LXe.  
These findings suggest that employing low-QE electrodes may help mitigate photoelectric-induced backgrounds, potentially improving the sensitivity of LXe TPCs in light dark matter searches.
\end{abstract}

\subjectindex{dark matter, liquid xenon, low background}

\maketitle

\section{Introduction}
\label{sec:intro}
Dual-phase time projection chambers (TPCs) that utilize both liquid and gaseous xenon are at the forefront of dark matter searches. These detectors have set the most stringent constraints on the interaction cross-section between Weakly Interacting Massive Particles (WIMPs) and ordinary matter~\cite{XENON:2023cxc, LZ:2022dm, PandaX:2024qfu}. By detecting primary scintillation (S1) and ionization signals via the proportional electroluminescence mechanism (S2), TPCs enable three-dimensional event reconstruction, which facilitates fiducialization of the active detector volume and suppression of radioactive backgrounds from detector materials and surface contaminants. Furthermore, the charge-to-light ratio (S2/S1) provides an effective means of discriminating nuclear recoil signals from electronic recoils based on differences in ionization density.

While the combination of S1 and S2 signals offers strong sensitivity to WIMPs above 5\,${\mathrm{GeV}/c^2}$, the sensitivity to low-mass WIMPs is limited due to the relatively low detection efficiency of S1 signals. Recent studies have demonstrated that liquid xenon (LXe) TPCs are highly sensitive to single ionization electrons as a result of proportional electroluminescence amplification in gaseous xenon (GXe), leading to a single-electron (SE) detection efficiency of nearly 100\,$\%$~\cite{XENON:2019gfn, XENON:2021nad, XENONnT_s2only, PandaX:2022xqx}. Leveraging this capability, SE-based analyses that do not require an S1 signal could significantly enhance the sensitivity to sub-GeV WIMPs. However, these analyses have been found to result in increased rates of low-energy electron backgrounds~\cite{XENON:2019gfn, XENON:2021nad, XENONnT_s2only, PandaX:2022xqx, LUXelectronbkgs}, originating from various sources, including impurities, delayed electron emission, and photoionization effects.

Among these background contributions, photoionization and the photoelectric effect induced by LXe scintillation photons on the electrode surfaces are of particular concern. Although these processes are recognized as potential contributions, their individual background rates remain poorly quantified due to the lack of dedicated measurements. This study aims to provide essential input for future Monte Carlo simulations by experimentally determining the quantum efficiency (QE) of candidate electrode materials for VUV photons in LXe. Quantifying the photoelectric contribution will help constrain this background component, thereby ultimately enhancing the sensitivity of upcoming S2-only dark matter searches, especially in the sub-GeV mass region.

In this study, we developed a dedicated experimental setup to measure the QE of various electrode materials for vacuum ultraviolet (VUV) photons in LXe and evaluated their potential for application in future LXe experiments such as DARWIN and XLZD~\cite{Aalbers:2016jon, XLZD:design}.

\section{Electrode Materials}
\label{sec:electrode}
In this study, stainless steel (SUS304), platinum (Pt), and magnesium fluoride (MgF$_{2}$)-coated aluminum (Al) were selected as candidate electrode materials, as summarized in Table~\ref{electrodes}. Stainless steel is widely used as an electrode material in current LXe experiments such as XENONnT~\cite{XENON:2023instrument_paper}, PandaX~\cite{PandaX:2024qfu}, and LZ~\cite{AKERIB2020163047}. It was chosen as the reference material in this study due to its well-established performance in these applications. For our measurements, an electropolished SUS304 plate with dimensions of 12\,mm\,$\times$\,12\,mm and a thickness of 2\,mm was used.

Platinum was selected as a candidate electrode material due to its high work function, the highest among common metals~\cite{work_function}, making it a promising choice for suppressing photoemission-induced backgrounds. A 100\,nm-thick Pt layer was deposited via sputtering onto a 12\,mm\,$\times$\,12\,mm quartz glass substrate with a thickness of 2\,mm. To enhance adhesion between the quartz glass and Pt, an intermediate chromium (Cr) layer with a thickness of 10\,nm was also deposited.

MgF$_{2}$-coated Al was chosen for its unique combination of reflective and insulating properties, which may contribute to a lower QE. This material is commonly used as a reflective flat mirror for VUV light~\cite{MgF2_mirror}, and the one produced by SIGMAKOKI (TFA-10S03-10) was used in this study. In this configuration, a 100\,nm-thick Al layer was first vapor-deposited onto a borosilicate glass substrate (12\,mm\,$\times$\,12\,mm, 3\,mm thick), followed by the vapor deposition of a 137\,nm-thick MgF$_{2}$ layer on top.

It is important to note that both the Pt and MgF$_{2}$ layers are sufficiently thick, ensuring that any photoelectrons generated in the underlying Cr or Al layers do not reach the surface and are therefore not extracted. This minimizes contributions to the measured QE from secondary effects.

\begin{table}[htbp]
\centering
\caption{Electrode materials investigated in this study}
\label{electrodes}
\begin{tabular}{|c||ccc|}
\hline
\textbf{Property} & \textbf{SUS304} & \textbf{Pt} & \textbf{MgF$_{2}$-coated Al} \\
\hline
Thickness & 2\,mm & 100\,nm & 137\,nm (MgF$_{2}$) \\
Intermediate layer & N/A & 10\,nm (Cr) & 100\,nm (Al) \\
Substrate & SUS304 & Quartz glass & Borosilicate glass \\
Deposition method & N/A & Sputtering & Vapor deposition \\
Surface treatment & Electropolished & N/A & N/A \\
\hline
\end{tabular}
\end{table}

\section{Experimental Setup and Operation}
\label{sec:setup_intro}
\subsection{Experimental Setup}
\label{sec:setup}

\begin{figure}[htbp]
    \centering
    \includegraphics[width=\textwidth]{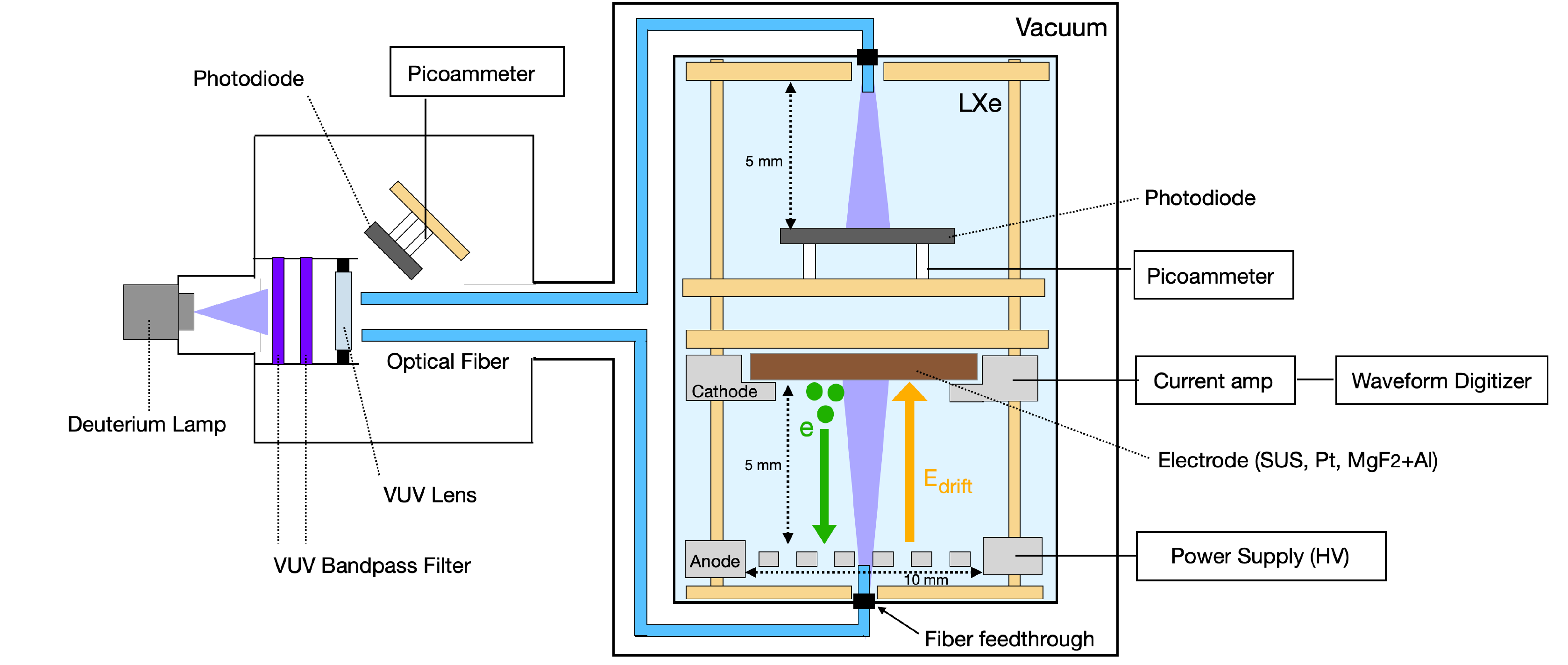}
    \caption{Experimental setup used to measure the QE of electrode materials.}
    \label{setup1}
\end{figure}

An experimental setup was developed at Nagoya University to measure the QE of electrode materials for VUV photons in LXe. 
The setup, illustrated in Fig.~\ref{setup1}, consisted of two primary components: one for determining the number of incident photons ($N_{\mathrm{ph}}$) and the other for estimating the number of electrons extracted from the cathode via the photoelectric effect and reaching the anode ($N_{\mathrm{e}}$). The QE is then calculated as
\begin{equation}
  \mathrm{QE} = \frac{N_{\mathrm{e}}}{N_{\mathrm{ph}}}
  \label{QE1}
\end{equation}

\begin{figure}[htbp]
    \centering
    \includegraphics[width=0.65\textwidth]{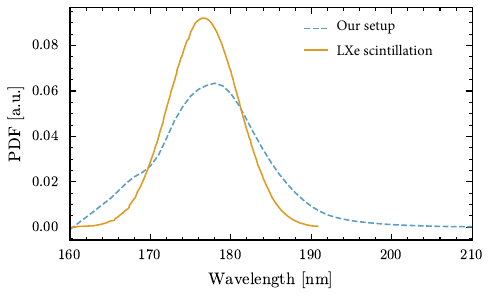}
    \caption{Expected deuterium lamp spectrum after applying the two VUV narrow bandpass filters (blue). The emission spectrum of LXe~\cite{lxe_wave} is also shown (orange).}
    \label{vuv_spectrum}
\end{figure}

A 110\,W deuterium lamp (Hamamatsu photonics, L15904) served as the VUV light source, producing continuous DC light with a broad spectrum ranging from 115 to 400\,nm. The lamp's MgF$_{2}$ window allows VUV photons to pass without significant attenuation. The spectral range was limited to the VUV region by two narrow bandpass filters (Pelham Research Optical, 172-NB-1D). Non-VUV components were mechanically blocked before entering the fiber by a combination of these filters and geometric alignment, as illustrated in Fig.~\ref{setup1}. A VUV lens (eSource Optics, CF 1220LCX) was then used to focus the filtered light beam onto the fibers. The resulting wavelength distribution, shown in Fig.~\ref{vuv_spectrum}, has a mean wavelength of 179.5\,nm with a standard deviation of 7.4\,nm, which closely matches the mean wavelength of the LXe scintillation photon ($\sim$174.8\,nm)~\cite{lxe_wave}. 
The VUV light was then transmitted through 2\,m long solarization-resistant quartz fibers (Molex, FDP 600660710).

The electrode sample was placed at the cathode, and a positive voltage was applied to the hexagonal mesh anode (2\,mm pitch and 0.1\,mm thickness) with a power supply (ISEG, NHR 4260r) to create a drift electric field between the electrodes. The VUV beam was focused onto the electrode surface with a narrow spot size that did not overlap with the anode mesh, ensuring that photoelectrons were generated only at the cathode. A 5\,mm thick PEEK spacer was placed between the cathode and the anode. Electrons emitted from the electrode surface via the photoelectric effect were drifted from the cathode to the anode by the applied electric field. The lateral size of the drift region is sufficiently large (approximately 10\,mm), making the effect of transverse diffusion in LXe (typically around 0.1\,mm) negligible and ensuring minimal loss of drifted electrons. The drift electric field between the cathode and the anode was varied from 100\,V/cm to 6\,kV/cm. The induced cathode current was recorded using a current amplifier (FEMTO, DDPCA-300) and a 14-bit waveform digitizer (CAEN, V1724G), from which $N_{\mathrm{e}}$ was determined.

To ensure accurate measurement of QE, $N_{\mathrm{ph}}$ was measured simultaneously with $N_{\mathrm{e}}$. The incident photon flux was determined with a photodiode (OPTO DIODE, AXUV100G) connected to a picoammeter (ADCMT, 8252 Digital Electrometer). The fiber-to-photodiode distance was fixed at 5~mm to ensure consistent light collection, matching the fiber-to-electrode distance used for $N_{\mathrm{e}}$ measurements. An additional photodiode (OPTO DIODE, AXUV100G) monitored the stability of the deuterium lamp during measurements. The instability of the light intensity was less than 1\,\%, indicating that the light source was sufficiently stable during the measurements.

\subsection{Operation}
\label{sec:operation}
As described in \cite{APRILE1994328, LUXelectronbkgs}, the QE is influenced by the band structure at the interface between the electrode material and the surrounding medium. To investigate this dependence, QE measurements were conducted under three different conditions: vacuum, GXe, and LXe. Additionally, the effective work function of metals decreases in the presence of an applied electric field due to the Schottky effect~\cite{APRILE1994328}, which predicts a reduction proportional to the square root of the electric field strength. This suggests that QE may also exhibit field dependence~\cite{LI2016160}.
To account for this, measurements were performed while varying the electric field strength.

The QE measurements in vacuum were performed at room temperature with the vacuum level maintained below $1\,\times\,10^{-3}~\mathrm{Pa}$.

\begin{figure}[htbp]
    \centering
    \includegraphics[width=0.8\textwidth]{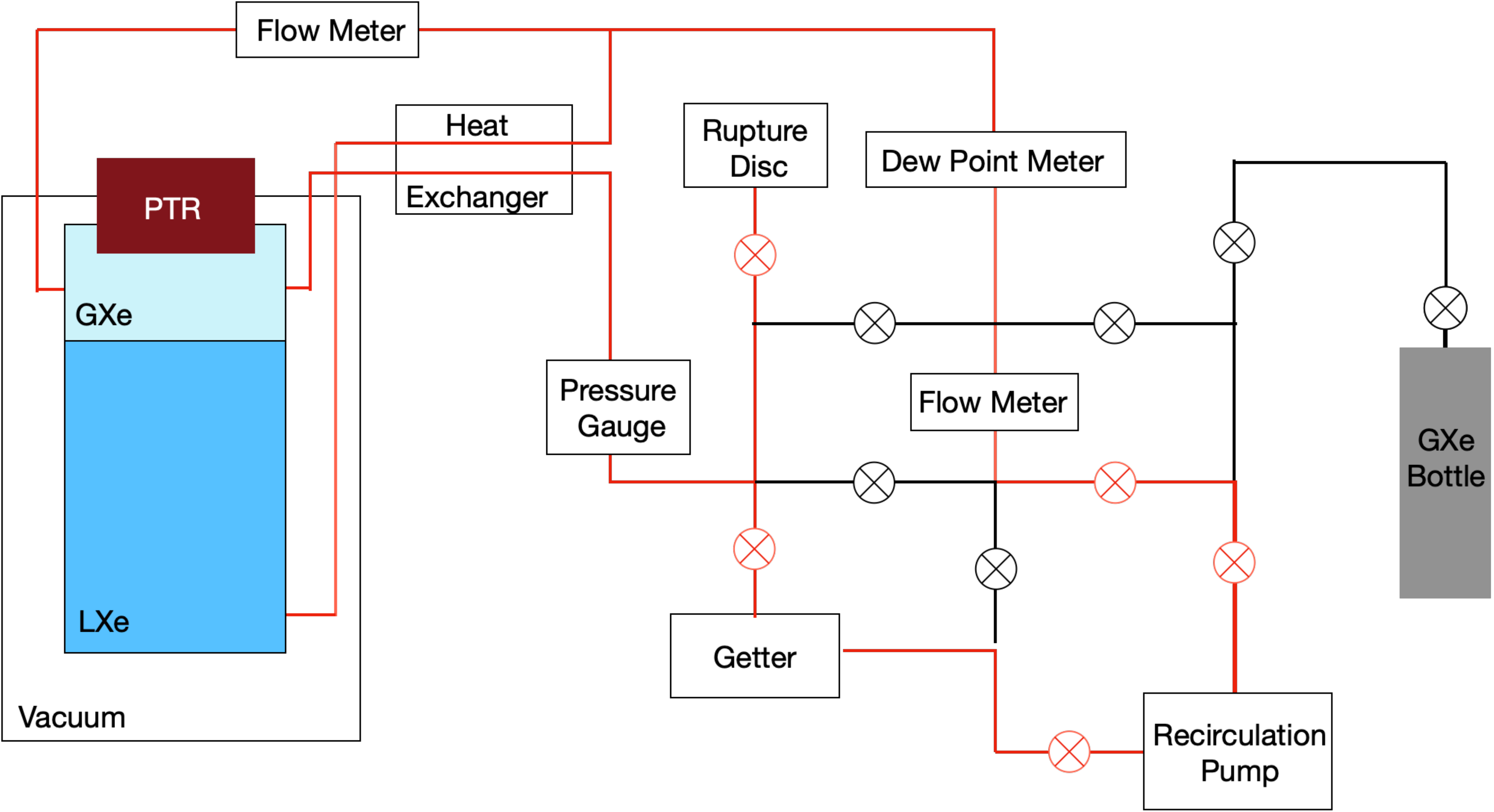}
    \caption{Xenon gas handling system used in GXe and LXe measurements. The line highlighted in red is the gas line used during the measurements in LXe.}
    \label{gas_system}
\end{figure}

For GXe measurements, xenon gas was introduced into the system at an absolute pressure of 150\,kPa. To ensure high purity and remove electronegative impurities that could capture free electrons, the gas was continuously circulated using a pump (Iwaki BA-330SN) and purified through a getter (SAES MicroTorr PS4-MT15-R-1). The gas handling system used for these measurements, applicable to both GXe and LXe configurations, is shown in Fig.~\ref{gas_system}. QE measurements in GXe were conducted at room temperature once the dew point reached $-$90$\,^\circ\,\mathrm{C}$, ensuring a high purity level. To evaluate the systematic uncertainty associated with noise instability, the measurements in vacuum and GXe were repeated three times under identical conditions.

For LXe measurements, GXe was liquefied using a pulse tube refrigerator (PTR, Iwatani PC-150), and approximately 55 standard liters of LXe were introduced into the detector, ensuring full immersion of the fiber tip delivering the VUV light. Due to the necessity of completing QE measurements for each sample within a single day, LXe measurements were typically carried out within an 8-hour window, during which temperature and pressure were continuously monitored. The representative temperature and pressure during these measurements were 173\,K and 160\,kPa (absolute pressure), with observed fluctuations of $\pm$3\,K and $\pm$20\,kPa, respectively. Measurements were repeated at both lower and higher points within these ranges, and the resulting variations were incorporated into the QE estimation as systematic uncertainties. These environmental fluctuations constituted the dominant contribution to the total systematic error.

\subsection{Measurement Procedure}
\label{sec:measurements}
As described in Sec.~\ref{sec:setup}, $N_{\mathrm{ph}}$ was measured using the photodiode signal and recorded with a picoammeter at a sampling rate of 1\,Hz for 100\,seconds under each applied electric field. The measured current was then converted to $N_{\mathrm{ph}}$ using the QE of the photodiode, which was calibrated by NIST. At the mean wavelength of 179.5\,nm, the QE was 83\,\%, with an associated uncertainty of approximately 5\,\%.
To correct for potential variations in light intensity between the two optical fibers, which ranged from approximately 5\,\% to 20\,\%, the fibers were exchanged at the fiber feedthrough of the LXe chamber, and $N_{\mathrm{ph}}$ was measured again under identical conditions. This fiber-swapping procedure cancels systematic differences in transmission efficiency between the two fibers. Since the fibers were securely mounted to maintain consistent alignment, uncertainties due to mechanical reproducibility or fiber bending are negligible. These effects are small compared to the dominant systematic uncertainty from temperature and pressure fluctuations, as discussed in Sec.~\ref{sec:operation}. The typical photocurrent was approximately 1\,nA.

$N_{\mathrm{e}}$ was determined by recording current signals from the cathode at a sampling rate of 1\,Hz with a waveform window of 100\,$\mu$s per event over 100\,seconds for each electric field. The typical cathode signal ranged from a few femtoamperes (fA) to picoamperes (pA), making it highly sensitive to noise. To mitigate this effect, baseline waveforms were recorded before and after lamp illumination for each electric field. The signal was then defined as the difference from the baseline during lamp illumination.

To account for systematic uncertainties due to temperature and pressure fluctuations as well as noise instability, the measurements were repeated three times as described in Sec.\ref{sec:operation}.

\section{Results and Discussion}
\label{sec:results}
\begin{figure}[htbp]
  \centering
  \begin{minipage}{0.60\columnwidth}
    \centering
    \includegraphics[width=\columnwidth]{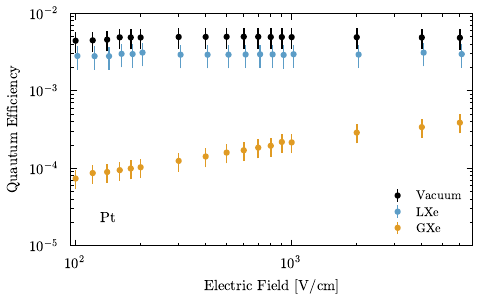}
    \label{QE_Pt}
  \end{minipage}
  \hspace{5mm}
  \begin{minipage}{0.60\columnwidth}
    \centering
    \includegraphics[width=\columnwidth]{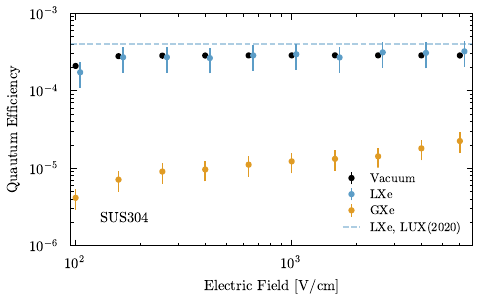}
    \label{QE_SUS}
  \end{minipage}
  \begin{minipage}{0.60\columnwidth}
    \centering
    \includegraphics[width=\columnwidth]{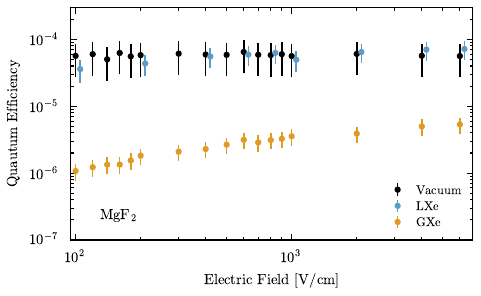}
    \label{QE_MgF2}
  \end{minipage}
  \caption{Quantum efficiency measured in vacuum, GXe and LXe for Pt (top), SUS304 (middle), and MgF$_{2}$ (bottom), respectively.}
  \label{QE_comparison}
\end{figure}

The QE measured in vacuum, GXe, and LXe for Pt, SUS304, and MgF$_2$-coated Al is shown in Fig.~\ref{QE_comparison}. The error bars at each data point reflect statistical uncertainties, differences in the baseline before and after measurements, and instabilities in temperature and pressure, with the latter being the dominant source of error, particularly in LXe measurements.

In this study, QE is defined as the number of electrons extracted from the cathode and reaching the anode per incident VUV photon. Given the current geometry, which features a small beam spot and a short drift length, the effect of transverse diffusion is negligible in both GXe and LXe. However, in GXe, a significant fraction of photoelectrons can be lost due to backscattering: electrons emitted from the electrode surface may elastically scatter off xenon atoms and return to the cathode before being collected. This process reduces the number of electrons contributing to the measured signal. 

The measured QE showed distinct behavior depending on the surrounding medium. In vacuum and LXe, QE exhibited little dependence on the applied electric field. In contrast, QE in GXe increased gradually with field strength. This trend is consistent with previous studies~\cite{Buzulutskov} and is attributed to the suppression of backscattering at higher fields: a stronger electric field pulls electrons away from the surface more effectively, reducing the chance that they are scattered back to the cathode by elastic collisions.

The variation in QE across vacuum, GXe, and LXe can be explained by two dominant factors: electron backscattering and modifications to the effective work function at the material–medium interface. In vacuum, where xenon atoms are absent, backscattering does not occur, resulting in the highest QE among the three media. In GXe, backscattering is prominent and reduces the effective QE, especially at low electric fields. In LXe, backscattering is strongly suppressed compared to GXe due to the medium’s much higher density and the correspondingly shorter electron mean free path. However, it may not be entirely negligible; a small fraction of emitted electrons can still be scattered back before escaping into the liquid.

Additionally, xenon’s negative electron affinity in LXe, approximately $-0.67$\,eV, reduces the effective work function at the electrode interface~\cite{APRILE1994328}, which tends to enhance the QE. In contrast, the electron affinity in GXe under our experimental conditions is close to zero and thus similar to that in vacuum, resulting in minimal reduction of the work function. The interplay between these effects, reduced backscattering and lower work function in LXe, can explain the observed differences in QE across the three media.

\begin{figure}[htbp]
    \centering
    \includegraphics[width=0.65\textwidth]{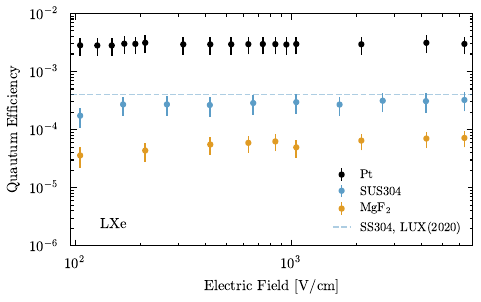}
    \caption{Comparison of quantum efficiency measured in LXe for Pt, SUS304, and MgF$_{2}$. QE of SS304 measured by the LUX experiment~\cite{LUXelectronbkgs} is also shown.}
    \label{fig:QE_all}
\end{figure}

The QE measured in LXe for Pt, SUS304, and MgF$_2$-coated Al is shown in Fig.~\ref{fig:QE_all}, with the corresponding values at an electric field of 6\,kV/cm summarized in Table~\ref{QE_table}. Among the tested materials, Pt exhibited the highest QE in LXe, while MgF$_2$-coated Al demonstrated a QE lower by a factor of approximately 4.4 compared to SUS304, a commonly used material in current direct dark matter detection experiments with LXe. The lower QE of SUS304 compared to Pt is likely due to the presence of a passive chromium oxide layer on its surface. In the LUX experiment, the QE of the gate electrode (SS304) for LXe scintillation photons has been evaluated to be approximately 3$\,\times\,10^{-4}$~\cite{LUXelectronbkgs}. Although a direct comparison cannot be made due to differences in the electrode geometry and electric field strengths used in the LUX detector, the results of this study are generally consistent with their result. 

\begin{table}[htbp]
\centering
\caption{Summary of QE measured at an electric field of 6~kV/cm in vacuum, GXe and LXe}
\smallskip
\begin{tabular}{|c||ccc|}
\hline
\textbf{Electrode Material} & \textbf{QE (Vacuum)} &  \textbf{QE (GXe)} & \textbf{QE (LXe)} \\
\hline
Pt & $(4.9\pm1.4)\times10^{-3}$ & $(3.9\pm1.1)\times10^{-4}$ & $(3.0\pm1.0)\times10^{-3}$ \\
SUS304 & $(2.9\pm0.2)\times10^{-4}$ & $(2.3\pm0.7)\times10^{-5}$ & $(3.2\pm1.1)\times10^{-4}$ \\
MgF$_{2}$-coated Al &  $(5.6\pm2.9)\times10^{-5}$ & $(5.3\pm1.5)\times10^{-6}$ & $(7.2\pm2.3)\times10^{-5}$ \\
\hline
\end{tabular}
\label{QE_table}
\end{table}

\section{Conclusion}
\label{sec:conclusion}
In this study, the QE of various electrode materials for VUV photons in LXe was investigated, focusing on their potential to suppress low-energy ionization backgrounds in dual-phase LXe TPCs for light dark matter searches. Measurements were performed for Pt, SUS304, and MgF$_2$-coated Al under vacuum, GXe, and LXe conditions.
Our results demonstrate that MgF$_2$-coated Al exhibits the lowest QE among the tested materials in LXe, measured to be $(7.2 \pm 2.3) \times 10^{-5}$, corresponding to a reduction by a factor of 4.4 compared to SUS304, a widely used electrode in current dark matter experiments. Pt exhibited the highest QE, while the relatively lower QE of SUS304 is likely attributed to its passive chromium oxide layer.
The reduction in QE observed with MgF$_2$-coated Al suggests its potential to mitigate electron emission induced by the photoelectric effect. However, its insulating nature may lead to surface charge accumulation and an increased risk of discharge. To address this issue, future developments should consider coatings with lower surface resistivity, such as antistatic materials, to suppress charge buildup while maintaining low QE.
These findings contribute to ongoing efforts to optimize electrode materials in LXe TPCs, with the potential to improve sensitivity to low-mass dark matter by reducing low-energy ionization backgrounds.

\section*{Acknowledgment}
We thank Masaki Yamashita (University of Tokyo) for providing the LXe setup used in this study. This work was supported by JSPS KAKENHI Grant Numbers 19H05805, 20H01931, 21H04466, and JST FOREST Program Grant Number JPMJFR212Q.

\vspace{0.2cm}

\bibliographystyle{apsrev4-2}
\bibliography{reference}

\noindent

\let\doi\relax

\end{document}